\documentclass[a4paper]{jpconf}
\usepackage{graphicx}
\usepackage[german,english]{babel}
\usepackage{upgreek}
\begin{document}
\title{Integration of Acoustic Neutrino Detection Methods into
  ANTARES\footnote{This work is supported by the German BMBF Grant
    No.~05~CN5WE1/7.}}

\author{K~Graf, G~Anton, J~H{\"o}{\ss}l, A~Kappes, U~F~Katz, R~Lahmann,
  C~Naumann and K~Salomon} \address{Physikalisches Institut,
  Friedrich-Alexander-Universit{\"a}t Erlangen-N{\"u}rnberg,
  Erwin-Rommel-Stra{\ss}e 1, D-91058 Erlangen, Germany}
\ead{graf@physik.uni-erlangen.de}

\begin{abstract}
  The ANTARES Neutrino Telescope \cite{antares_proposal} is a water
  Cherenkov detector currently under construction in the Mediterranean
  Sea. It is also designed to serve as a platform for investigations
  of the deep-sea environment. In this context, the ANTARES group at
  the University of Erlangen will integrate acoustic sensors within
  the infrastructure of the experiment. With this dedicated setup,
  tests of acoustic particle detection methods and deep-sea acoustic
  background studies shall be performed. The aim of this project is to
  evaluate the feasibility of a future acoustic neutrino telescope in
  the deep sea operating in the ultra-high energy regime. In these
  proceedings, the implementation of the project is described in the
  context of the premises and challenges set by the physics of
  acoustic particle detection and the integration into an existing
  infrastructure.
\end{abstract}

\section{Introduction}

Towards the detection of neutrinos with energies exceeding 100\,PeV,
the use of acoustic pressure waves produced in neutrino-induced
cascades is a promising approach. One advantage of acoustic waves is
the absorption length of the order of 1\,km for the peak spectral
density of the generated sound waves around 20\,kHz
\cite{fisher_jasa}.  To investigate the feasibility of building a
detector in the deep sea based on this method, it is necessary to
understand the acoustic background conditions and characteristics of
transient noise sources in detail on a long time scale.  Especially
the rate and correlation length of neutrino-like acoustic background
events is not known and yet is a prerequisite for the estimation of
the sensitivity of such a detector.  The aim of the project described
here is to measure the acoustic conditions of the deep-sea environment
at the ANTARES site with a dedicated array of custom-designed acoustic
sensors at different distances.

Towards this goal several additional basic detector elements ({\it
  storeys}, cf.~Sec.~\ref{cap_antares}) of the ANTARES neutrino
telescope will be equipped with acoustic sensors. On these storeys the
sensors will substitute the optical sensors (photo-multiplier tubes,
{\it PMTs}) used for Cherenkov detection of neutrinos.  Several
components of the ANTARES infrastructure have to be modified or
substituted to integrate the acoustic sensors into the ANTARES data
acquisition ({\it DAQ}) scheme. In these proceedings the changes will
be described, and an overview of the project will be given (cf.\ also
\cite{lahmann_arena05}).

\section{The ANTARES detector}\label{cap_antares}
The ANTARES detector \cite{antares_proposal} is currently installed in
the Mediterranean Sea, 40\,km off the coast of Toulon (southern
France) in a water depth of up to 2500\,m. Its completion is foreseen
in 2007; the neutrino telescope will then consist of 12 vertical
structures ({\it detection lines}) with a total of 900 optical sensors
for the detection of Cherenkov light of neutrino interaction
secondaries. Each line is fixed to the sea bed by an anchor and held
vertically by a buoy on the top. A detection line has a total height
of 480\,m and comprises 25 storeys spaced evenly within the
instrumented height of 350\,m. The 12 lines are placed on the sea
floor in an octagonal shape with 4 lines in a quadratical layout in
the centre.  The distance between two neighbouring lines ranges from
60\,m to 80\,m, covering a total area of approx.\
180\,$\times\,$180\,m{$^2$} on the sea-floor.  An extra 13th line \--
the Instrumentation Line ({\it IL}) \-- will be equipped with sensors
to monitor environmental parameters and with devices to calibrate the
detector. The detector is connected to the on-shore control room via
deep-sea cables providing electrical power and data transmission. At
the writing of these proceedings, two detection lines and a progenitor
of the IL are installed at the site and operated successfully.

\begin{figure}[h]
\centering
\includegraphics[width=15cm]{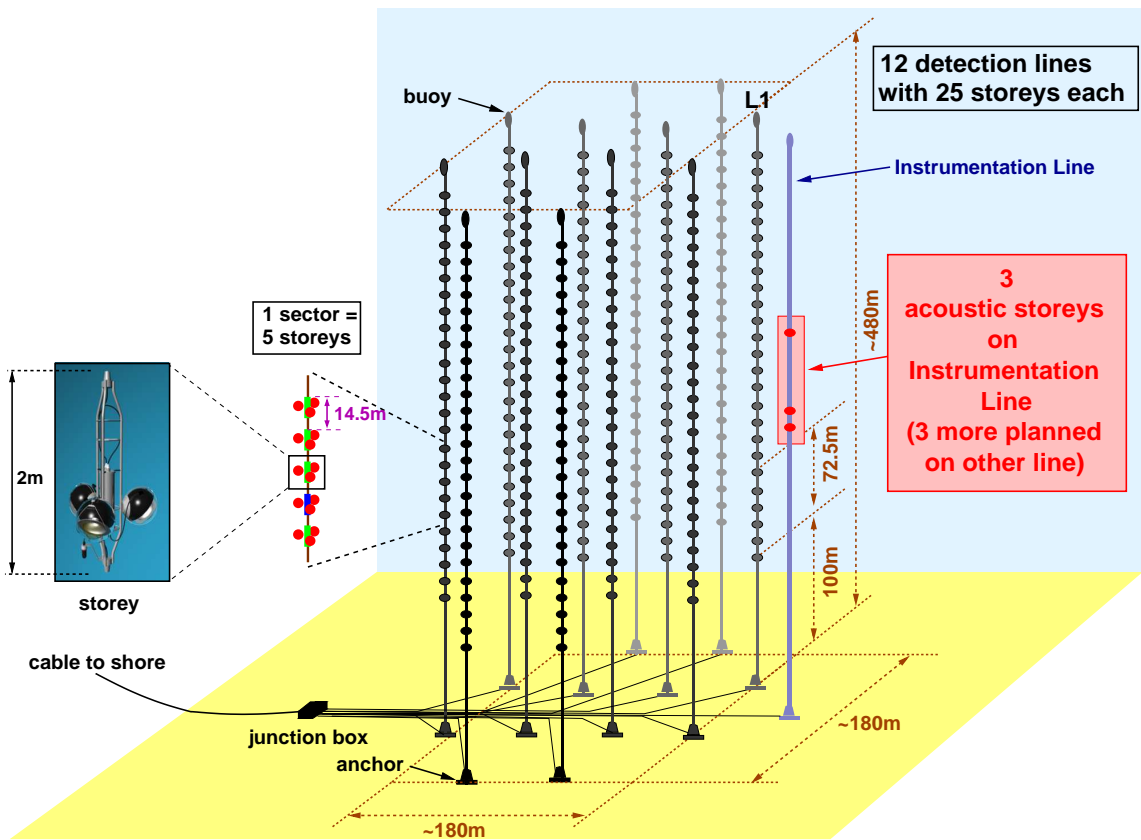}\hspace{2pc}%
\caption{Sketch of the ANTARES detector with the addition for acoustic
  studies. For further description see
  text.\label{fig_antares_scheme}}
\end{figure}

\begin{figure}[h]
\includegraphics[height=6cm]{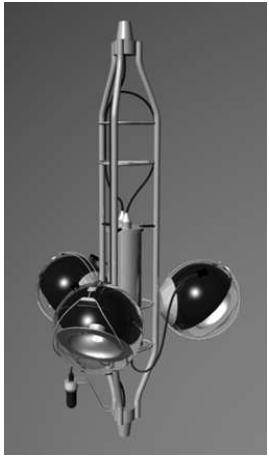}\hspace{2pc}%
\begin{minipage}[b][6cm]{10cm}
  \caption{An ANTARES storey equipped with three optical sensors in
    glass spheres and a container for electronics (Titanium cylinder
    in the middle) attached to a Titanium support structure. Visible
    at the lower left is a hydrophone used for the determination of
    the storey position.\label{fig_antares_storey_opt}}
\end{minipage}
\end{figure}

A sketch of the complete ANTARES detector with the acoustic addition
described in Sec.~\ref{sec_acoustics} is shown in
Fig.~\ref{fig_antares_scheme}.
\\
An optical storey (cf.~Fig.~\ref{fig_antares_storey_opt}) constitutes
the basic detection element of the ANTARES detector and consists of
three Optical Modules ({\it OMs}) (optical sensors in a
pressure-resistant glass housing), a Local Control Module ({\it LCM})
(for data acquisition-, control- and monitoring hardware) and miscellaneous auxiliary devices on a mechanical support frame.

\section{Aim of the Acoustic Studies in ANTARES}
Neutrinos interacting in water deposit energy in a cascade of
secondary particles. According to the thermo-acoustic model developed
by Askarian \cite{askarian_nim}, the fast deposition and slow
dissipation of the energy in the water in form of heat leads to a
bipolar pressure pulse ({\em BIP}) \cite{graf_arena05}. The main
frequency range of the generated sound waves spans from 1 to 100\,kHz
with a maximum spectral density around 20\,kHz.  It propagates through
the medium in a flat disk shape perpendicular to the main axis of the
cascade.  

Given the expected low flux of neutrinos with energies in excess of
100\,PeV, a potential acoustic neutrino telescope not only must have
very large dimensions, but must be operated basically background-free.
It is therefore of great importance to understand the acoustic
background due to BIPs from noise sources in the deep sea, which could
mimic neutrino signatures in a future detector. The measurement of the
rate of BIP signals as a function of pressure amplitude and volume is
therefore decisive to assess the feasibility of acoustic neutrino
telescopes and to determine a realistic lower bound on the neutrino
energy that is detectable. 

The acoustic detection equipment in ANTARES is designed to perform
such background measurements over a period of several years in a
realistic arrangement of acoustic sensors. Simulations \cite{karg_phd}
suggest that an acoustic detector will not gain efficiency
significantly when instrumented with more than 200 acoustic clusters
(each consisting of several hydrophones, similar to the acoustic
storeys in ANTARES) per km$^3$. The resulting average distance between
acoustic clusters is on the order of 200\,m.  This is comparable to
the largest distance between acoustic storeys in ANTARES, as will be
discussed in the next section.

\section{The Acoustic Setup}\label{sec_acoustics}
Acoustic sensors will be integrated at three storeys on the IL, which
will be installed at the ANTARES site in mid-2007. Each storey will be
equipped with 6 sensors. The vertical distance for these storeys will
be approx.\ 15\,m and 100\,m, resulting in \--- together with the
sensor spacing of approx.\ 1\,m within the storey \--- three different
length scales for the investigation of acoustic background sources.
Three additional acoustic storeys are planned on one further detection
line at a horizontal distance exceeding 100\,m.

\subsection{The Acoustic Sensors}

For the integration of acoustic sensors into the ANTARES experiment,
the data acquisition system \cite{antares_daq} has to be modified.
This is done under the premise of preventing any interference with the
optical data taking. To optimise resources and to make use of the
well-tested existing system wherever feasible, as few changes as
possible to the ANTARES design are targeted.

\begin{figure}[h]
\centering
\includegraphics[height=6cm]{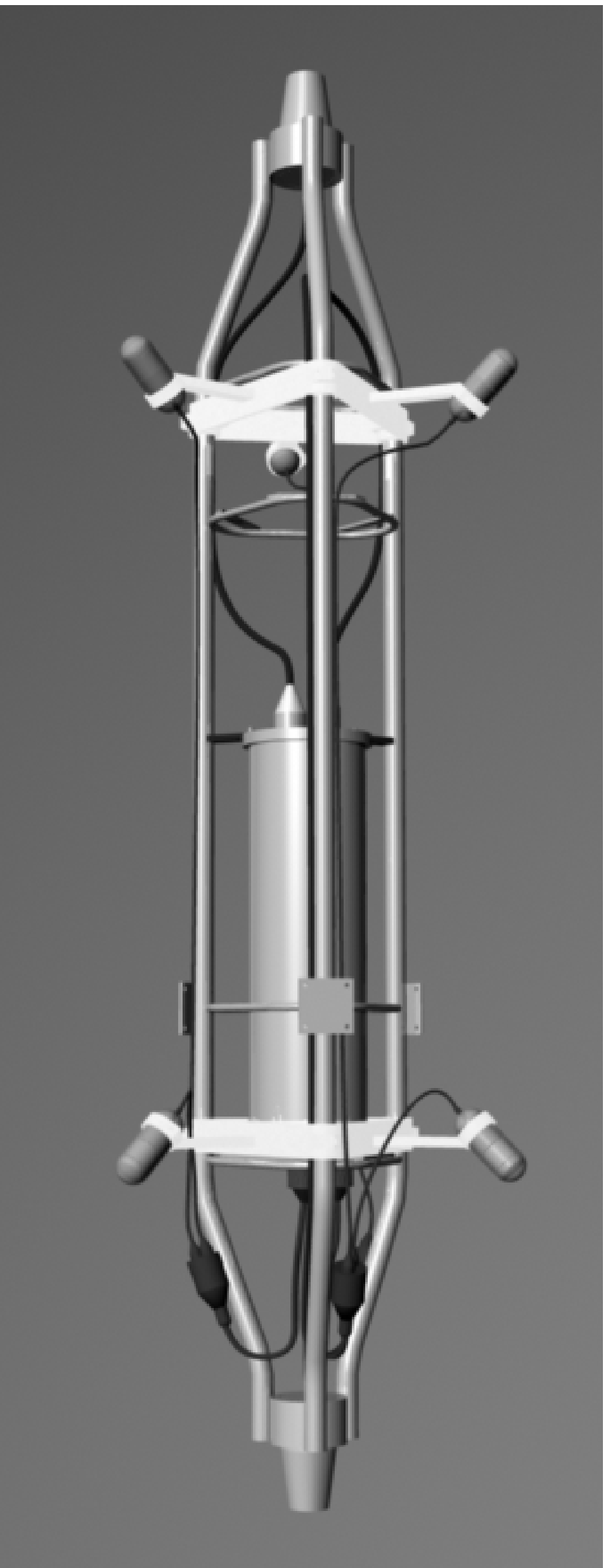}\hspace{2pc}%
\includegraphics[height=6cm]{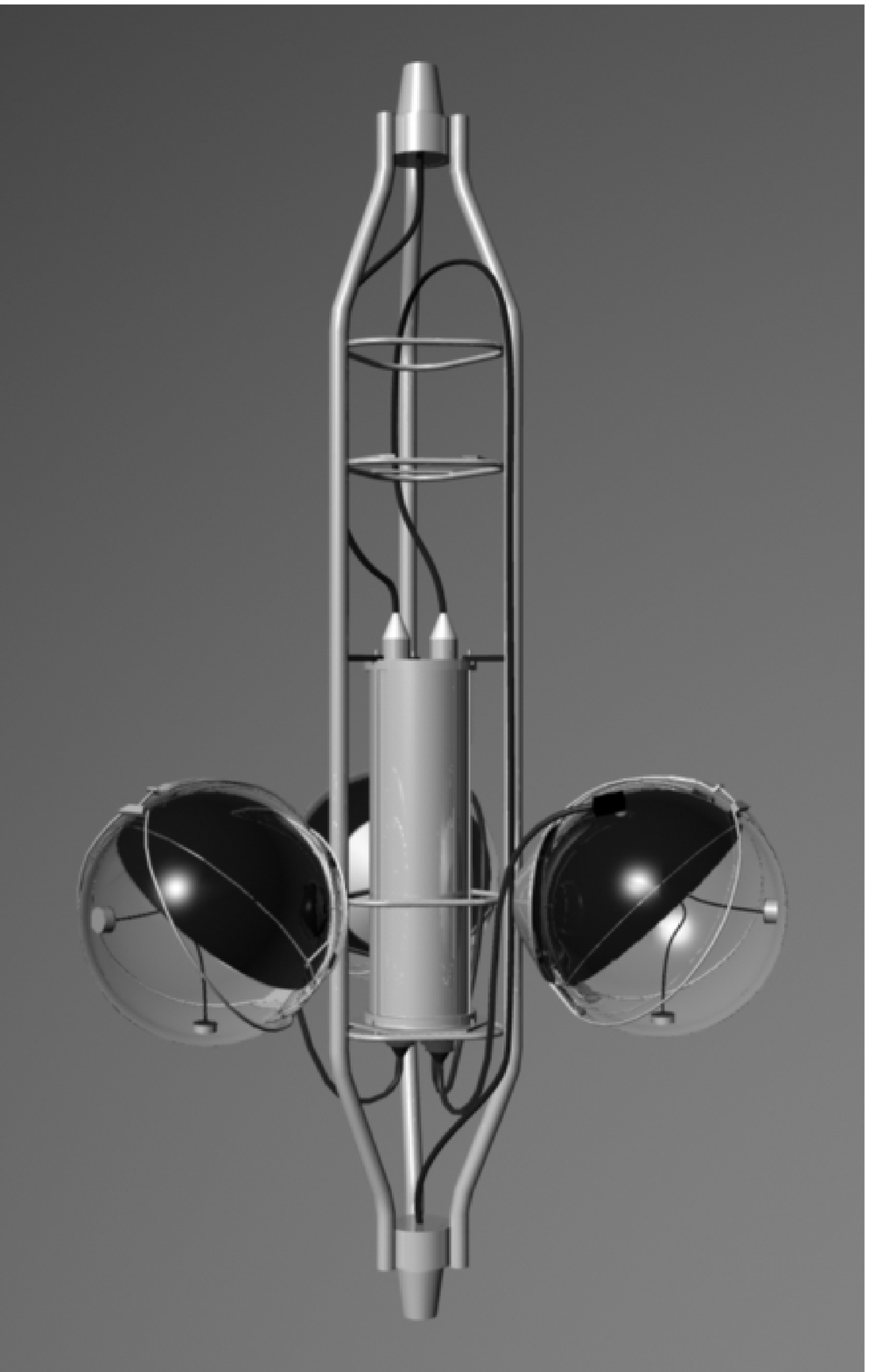}\hspace{2pc}%
\begin{minipage}[b]{8cm}
  \caption{Artist's views of acoustic storeys resulting from the
    replacement of OMs by six hydrophones (left) or three acoustic
    modules housing two piezo sensors each
    (right).\label{fig_antares_storey_acou}}
\end{minipage}
\end{figure}

Major changes affect the storey, where the OMs are replaced by
custom-designed acoustic sensors: hydrophones or so called acoustic
modules ({\it AMs}) (cf.~Fig.~\ref{fig_antares_storey_acou}). These
sensors are based on piezo-electrical ceramics that convert pressure
waves into voltage signals, which are then amplified for read out
\cite{hoessl_app}. The ceramics and amplifiers are coated in polymer
plastics in the case of the hydrophones. For the AMs they are glued to
the inside of a water-tight sphere. The latter non-conventional
design was inspired by the idea to take the deep-sea water pressure
from the piezo ceramics and electronics.  The three acoustic storeys
on the IL will house hydrophones only, whereas at least one of storeys
planned in addition will house AMs. 

\begin{figure}[h]
\centering
\includegraphics[height=7cm]{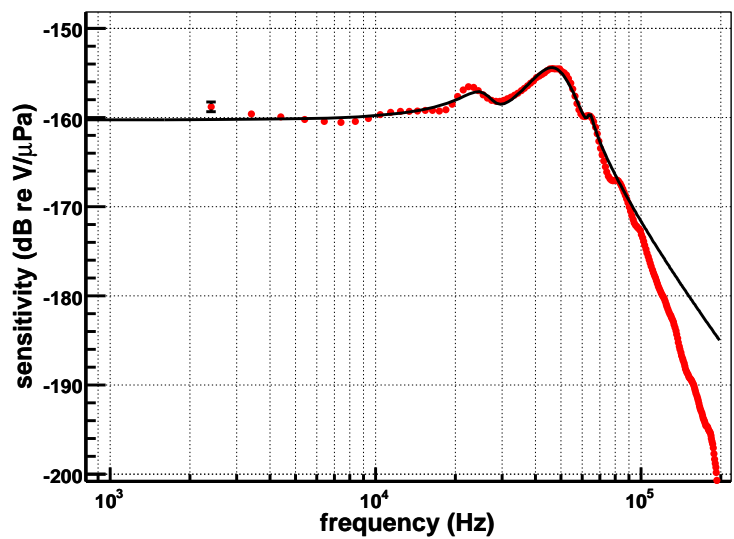}\hspace{2pc}%
\caption{Sensitivity plot of a prototype hydrophone with measured data
  points and a model fit \cite{hoessl_app} (solid line) in the
  frequency range from 1 to 250\,kHz on a logarithmic scale.  The
  model does not include the frequency response of the amplifier in
  the sensor, which is responsible for the discrepancy at higher
  frequencies.\label{fig_hydrophone_sensitivity}}
\end{figure}

All acoustic sensors are tuned to be sensitive over the whole
frequency range of interest (typically around
-145\,dB\,re.\,1V/$\upmu$Pa) and to have a low noise level
\cite{naumann_arena05}. The sensitivity of a prototype hydrophone with
approx.\ 10\,dB less gain than the final sensor design is shown in
Fig.~\ref{fig_hydrophone_sensitivity}.  Artist's views of the storeys
with different acoustic sensors are shown in
Fig.~\ref{fig_antares_storey_acou}.

\subsection{The Acoustic Data Acquisition}

In the ANTARES DAQ scheme the digitisation is conducted within the LCM
on each storey by several custom-designed electronics boards, which
send as much unfiltered data as possible to shore. With its capability
of timing resolutions on a nanosecond-scale and synchronous transmission
of several MByte per second and storey, it is perfectly suited for the
acquisition of acoustic data. Additionally, the ANTARES infrastructure
provides measurements of the position of the storey with a precision
better than 10\,cm and of miscellaneous environmental parameters
(e.g.\ the speed of sound, temperature and sea current).  For the
digitisation of the acoustic signals and feeding into the ANTARES data
stream the so-called acoustics digitisation board ({\it
  AcouADC-board}) was designed. There are a total of three such boards
per storey receiving the data of two sensors each.

\subsection{The AcouADC-Board}

The AcouADC-board consists of an analogue and a digital part. The
analogue part amplifies the voltage signals coming from the acoustic
sensors by adjustable factors between 1 and 512 and filters the
resulting signal.  The system has low noise and is designed to be \---
together with the sensors \--- sensitive to the acoustic background of
the deep sea over a wide frequency-range (approx. 1 to 100\,kHz).

\begin{figure}[h]
\begin{minipage}[c]{8.0cm}
\centering
\includegraphics[height=5.7cm]{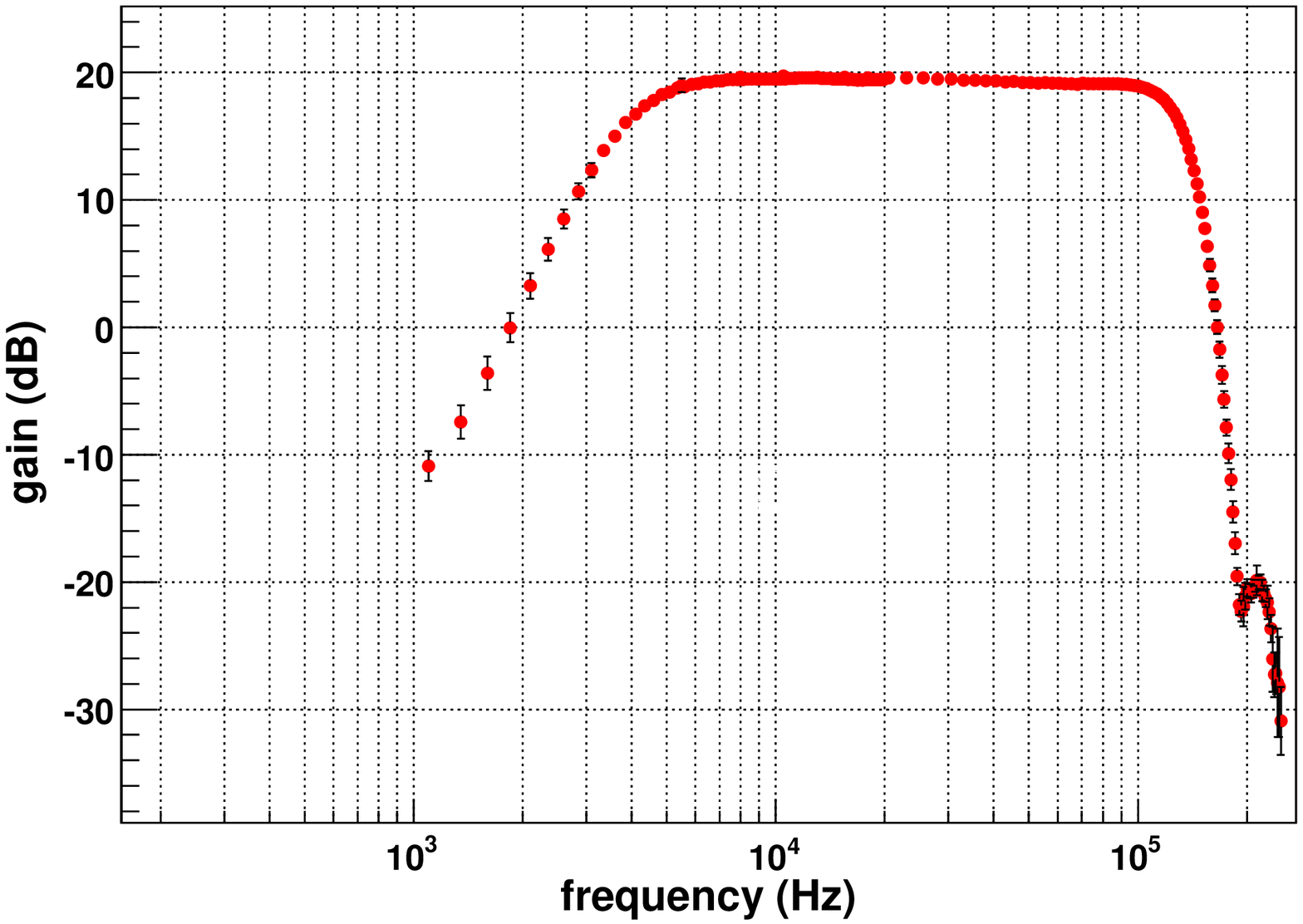}\hspace{2pc}%
\end{minipage}
\begin{minipage}[c]{8.0cm}
\centering
\includegraphics[height=5.7cm]{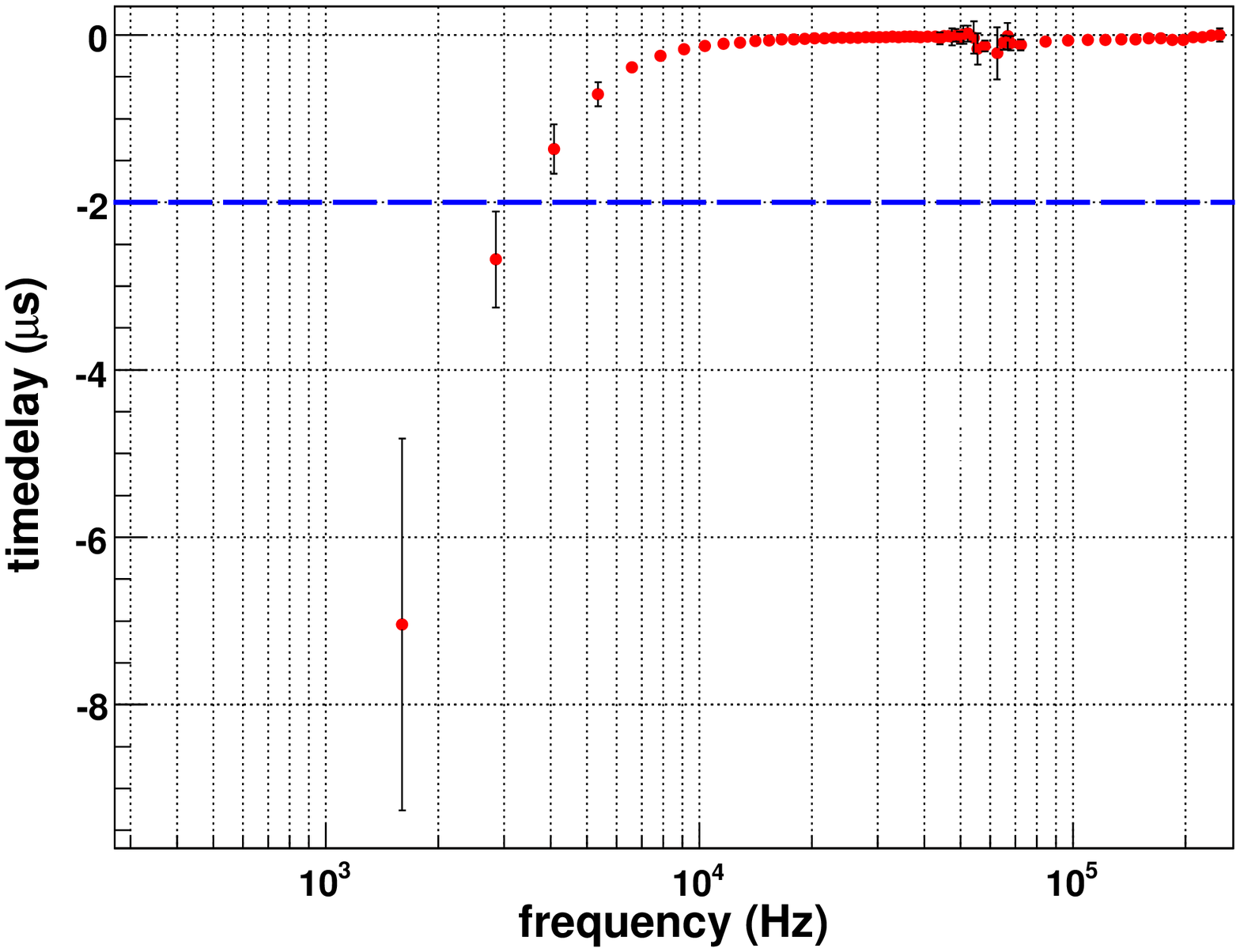}\hspace{2pc}%
\end{minipage}
\caption{Measured filter characteristics of the analogue part of a
  prototype AcouADC-board in the frequency range from 1 to 250\,kHz on
  a logarithmic scale. The left figure shows the amplitude gain in dB,
  the right one the time delay in $\upmu$s. The dashed line in the
  right-hand plot denotes the electronics digitisation time of
  2\,$\upmu$s, showing the characteristic time
  scale.\label{fig_frequency_response}}
\end{figure}

Figure~\ref{fig_frequency_response} shows the measured filter
characteristics of the analogue part of a prototype AcouADC-board for
a gain of 10 in amplitude (20\,dB). In the left hand plot the
frequency response of the integrated bandpass is observable. This
filter suppresses frequencies below approx.\ 4\,kHz and above approx.\
130\,kHz. The high-pass part cuts into the trailing edge of the low
frequency noise of the deep-sea acoustic background \cite{urick} and
thus protects the system from saturation. The low-pass part
efficiently suppresses frequencies above the Nyquist frequency of
250\,kHz for the digitisation frequency of 500\,kHz (see below). For
frequencies above roughly 10\,kHz, the time delay of signals passing
through the analogue part (right plot of
Fig.~\ref{fig_frequency_response}) is negligible on the digitisation
time scale of 2\,$\upmu$s.
\\
By the calibration of the whole data taking chain \--- the transfer
function of the system \--- it is possible to reconstruct the acoustic
signal from the recorded one with high precision within the sensitive
frequency range of the setup. The dynamic range achieved is from the
order of 1\,mPa to the order of 10\,Pa in rms over the frequency range
from 1 to 100\,kHz. 

This allows for studying both the acoustic background in the deep sea
under all prevailing conditions \cite{urick} and BIPs from noise
sources mimicking neutrino signatures of energies exceeding
e.g.~GZK-neutrinos by several orders of magnitude.

The digital part digitises and processes the acoustic data. It is
designed to be highly flexible by employing a micro controller ({\it
  $\mu$C}) and a field programmable gate array ({\it FPGA}) as data
processor. The $\upmu$C can be controlled from the on-shore control
room and is used to adjust settings of the analogue part and the data
processing.  Also, the complete programming of the AcouADC-board can be
updated in situ.  The FPGA reads the digitised acoustic data from the
analogue to digital converter ({\it ADC}) which has a 16-bit
resolution and a maximum sampling rate of 500\,k\,Samples per second
({\it kSPS}). The data is processed, e.g.~down-sampled to reduce data
traffic, and is read out by the ANTARES DAQ system which handles the
transmission to the control room.

The maximum data rate per storey is limited to 20\,Mbit per second by
the electronics sending the data to shore, limiting the maximum sample
rate of acoustic data per storey to 1.25\,MSPS.  Therefore the digital
data is down-sampled to 200\,kSPS at the AcouADC-board, to allow for
continuous and synchronous read-out of all six sensors of each storey.
However, the bandwidth can be distributed freely between the sensors,
so transmission of the acoustic data is also possible for two sensors
at each storey at full sampling-rate.

On-shore a dedicated PC-cluster will be set up to process and store
the acoustic data arriving from the storeys and to control the off-shore
DAQ. On this cluster, different data filtering schemes and triggers
will be implemented, as well as raw data stored.

\section{Summary and Outlook}

A dedicated array of acoustic sensors will be installed in the
deep-sea environment of the ANTARES site in 2007. The aim is to study
the feasibility of a future neutrino detector in water employing the
acoustic detection method. The technical realisation of the project is
well advanced with prototypes of each component ready and tested. The
setup will provide the possibility to study the acoustic background
noise and signals of the deep sea on a long time scale in the
frequency range of interest between 1 and 100\,kHz. The main goal is
to measure the rate of correlated neutrino-like background events on
different length scales, which is decisive for assessing the
sensitivity of a future acoustic detector for ultra-high-energy
neutrinos.

\smallskip
\end{document}